\documentclass[aps, prl, twocolumn, superscriptaddress, showpacs]{revtex4}
\usepackage{amsfonts}
\usepackage{bbm}

\usepackage{graphicx}
\begin{document}

\title{Spin superconductor in ferromagnetic graphene}

\author{Qing-feng Sun}
\email{sunqf@aphy.iphy.ac.cn}
\affiliation{Institute of Physics,
Chinese Academy of Sciences, Beijing 100190, China}
\author{Zhao-tan Jiang}
\affiliation{Department of Physics, Beijing Institute of Technology,
Beijing 100081, China}
\author{Yue Yu}
\affiliation{Institute of Theoretical Physics, Chinese Academy of
Sciences, Beijing 100190, China}
\author{X. C. Xie}
\affiliation{International Center for Quantum Materials, Peking
University, Beijing 100871, China}
\affiliation{Institute of
Physics, Chinese Academy of Sciences, Beijing 100190, China}
\affiliation{Department of Physics, Oklahoma State University,
Stillwater, Oklahoma 74078}

\date{\today}

\begin{abstract}
We show a spin superconductor (SSC) in ferromagnetic graphene as
the counterpart to the charge superconductor, in which a
spin-polarized electron-hole pair plays the role of the spin $2
(\hbar/2)$ `Cooper pair' with a neutral charge. We present a
BCS-type theory for the SSC. With the `London-type equations' of
the super-spin-current density, we show the existence of an
electric `Meissner effect' against a spatial varying electric
field. We further study a SSC/normal conductor/SSC junction and
predict a spin-current Josephson effect.
\end{abstract}


\maketitle

The superconductivity was discovered about a century ago.\cite{sc}
Since then, it has been one of the central subjects in
physics.\cite{sc} Many fascinating properties of superconductors,
such as zero resistance,\cite{sc} the Meissner effect,\cite{meiss}
and the Josephson effect,\cite{jose} have many applications
nowadays. On the other hand, the potential application of the spin
degrees of freedom of an electron, the field of spintronics, is
still rapidly developing and is emerging as a major field in
condensed matter physics.\cite{spintronics}

The key physics of superconductivity was well understood in the BCS
theory:\cite{BCS} Electrons in a solid state system may have a net
weak attraction so that they form Cooper pairs which can then
condense into the BCS ground state. The simplest $s$-wave Cooper
pairs are of electric charge $2e$ and spin singlet. A dual of
superconductor is the so-called exciton condensate in which a Cooper
pair-like object is a particle-hole pair which is charge-neutral
while its spin may either be singlet or triplet. We name a
spin-triplet exciton condensate as the spin superconductor (SSC).
The exciton condensates can exist in many physical
systems.\cite{exciton1,exciton2} However, a general drawback of the
exciton condensate is its instability because of electron-hole (e-h)
recombination that lowers the total energy of the system (see
Fig.1(a)). The typical lifetime of an exciton is restricted from
pico-second to nano-second, and at most limited to the micro-second
range.\cite{timeps,timens,timereview} It is too short for many
meaningful applications. Thus, finding a long-lived exciton gas
becomes an important task.

Recently, graphene, a single-layer hexagonal lattice of carbon
atoms, has been successfully fabricated.\cite{ref1a,ref1b,ref3,ref4}
The unique structure of graphene leads to many peculiar properties,
e.g., the relativistic-like quasi-particles
spectrum.\cite{ref3,ref4} For graphene, the charge carriers are
usually spin-unpolarized. However, if graphene is growing on a
ferromagnetic (FM) material\cite{ref5,ref6,ref7} or is under an
external magnetic field\cite{a2ref1}, a spin split $M$ can be
induced. Then the carriers are spin-polarized and the Dirac points
with different spins are split. When the Fermi level lies in between
the spin-resolved Dirac points (see Fig. \ref{fig:Fig.1}(b)), the
spin-up carriers are electron-like while the spin-down ones are
hole-like. These positive and negative carriers attract and form e-h
pairs that are stable against the e-h recombination due to the
Coulomb interaction. This is because the filled electron-like states
are now below the hole-like states, as shown in Fig.
\ref{fig:Fig.1}(b), unlike in conventional exciton systems in
semiconductors (Fig. \ref{fig:Fig.1}(a)) where the electron states
are above the hole states. If a carrier jumps from the electron-like
state to the hole-like one, the total energy of the system rises.
This prevents the e-h recombination and means the e-h pairs in FM
graphene is stable and can exist indefinitely in princeple.
Therefore, this e-h pair gas can condense. In this work, we show
that this condensate is a SSC and the spin current is
dissipationless. We derive the London-type equations of the
super-spin-current and find an electric `Meissner effect' against
the spatial variation of an electric field.

We consider an interacting electron system in graphene with the
Hamiltonian $H=H_0+U_C$ where $H_0$ is the free Dirac fermion
Hamiltonian and $U_C$ the electron-electron (e-e) Coulomb
interaction:
\begin{eqnarray}
  H_0&=&\sum\limits_{{\bf k},\sigma}
  \Psi^{\dagger}_{{\bf k}\sigma}
    \left(\begin{array}{ll} -\sigma M & v_F(k_x -i k_y) \\ v_F(k_x +i k_y) & -\sigma M \end{array}\right)
    \Psi_{{\bf k}\sigma}
 , \nonumber \\
 U_C &=& \sum\limits_{s,s';i,j;\sigma,\sigma'} U^{ss'}_{ij} n^s_{i\sigma}
 n^{s'}_{j\sigma'},
\end{eqnarray}
where $\Psi_{{\bf k}\sigma}=( a_{{\bf k}\sigma}, b_{{\bf k}\sigma}
)^T$, $s_{{\bf k}\sigma}$ $(s=a,b)$ are the Fourier components of
the electron annihilation operators $s_{i\sigma}$ at sites $i$ for
the sublattices $s$, and $n^{s}_{i\sigma}=s^{\dagger}_{i\sigma}
s_{i\sigma}$ are the local electron number operators. ${\bf
k}=(k_x,k_y)$ is the momentum, $\sigma=(\uparrow,\downarrow)$
represents the spin, $M$ is the FM exchange split energy,
$U^{ss'}_{ij}$ is the e-e Coulomb potential, and $v_F =3ta_0/2$ with
the nearest hopping energy $t$ and the carbon-carbon distance $a_0$.
Here we have ignored the valley degree of freedom, because the two
valleys are degenerate and the inter-valley coupling is normally
very weak due to the two valleys being well separated in k-space.
Hereafter we also set the Fermi energy $E_F$ at zero. By taking a
unitary transformation: $a_{{\bf k}\sigma}=\sum_{\tau}\tau c^*
\alpha_{{\bf k}\tau\sigma}$ and $b_{{\bf k}\sigma}= \sum_{\tau} c
\alpha_{{\bf k}\tau\sigma}$ with the pseudo spin index $\tau=\pm$,
$c=e^{i\theta/2}/\sqrt{2}$ and $\theta=\tan^{-1}(k_y/k_x)$, the free
Hamiltonian $H_0$ can be diagonalized $H_0= \sum_{{\bf k}, \tau,
\sigma} \epsilon_{\tau \sigma}\alpha_{{\bf
k}\tau\sigma}\alpha^{\dagger}_{{\bf k}\tau\sigma} $, where
$\epsilon_{\tau\sigma}=-\sigma M + \tau v_F k$ are four energy bands
(see the blue curves in Fig.1d) because the spin-degeneracy is
lifted now. While $E_F=0$, $\epsilon_{-\uparrow}$ and
$\epsilon_{+\downarrow}$ are high-energy bands. In the following, we
focus on the low energy part and only two bands
$\epsilon_{+\uparrow}$ and $\epsilon_{-\downarrow}$ are involved.
Here the band $\epsilon_{+\uparrow}$ is electron-like, while the
band $\epsilon_{-\downarrow}$ is hole-like and the annihilation
operator $\alpha_{{\bf k}-\downarrow}$ also means to create a
spin-up hole. Thus, we define operators $\alpha_{{\bf k}e\uparrow}
=\alpha_{{\bf k}+\uparrow}$ and $\alpha^{\dagger}_{{\bf k}h\uparrow}
=\alpha_{{\bf k}-\downarrow}$. The Hamiltonian $H_0$ can then be
written as:
\begin{eqnarray}
  H_0&=& \sum\limits_{\bf k}( \alpha^{\dagger}_{{\bf k}e\uparrow}, \alpha_{{\bf k}h\uparrow}
    )
    \left(\begin{array}{ll} \epsilon_{+\uparrow} & 0 \\ 0 & \epsilon_{-\downarrow} \end{array}\right)
 \left( \begin{array}{l}
 \alpha_{{\bf k}e\uparrow} \\ \alpha^{\dagger}_{{\bf k}h\uparrow}
 \end{array}
 \right).
\end{eqnarray}
For the e-e interaction $U_C$, we also focus on the two low-energy
bands which are given by the terms $\alpha^{\dagger}_{{\bf k}-{\bf
q} ,e\uparrow}\alpha_{{\bf k}e\uparrow}  \alpha_{{\bf k}'+{\bf q} ,
h\uparrow} \alpha^{\dagger}_{{\bf k}' h\uparrow}$. Furthermore, we
keep only the terms whose momenta satisfy ${\bf k}-{\bf q} ={\bf
k}'$, giving rise to the zero momentum e-h pair that is
energetically favorable. Under these approximations, the interaction
$U_C$ reduces to the attraction between electrons and holes
\begin{eqnarray}
  U_{C}= -\sum\limits_{{\bf k}, {\bf k}'} U_{{\bf k}{\bf k}'}
  \alpha^{\dagger}_{{\bf k}'e\uparrow}   \alpha^{\dagger}_{{\bf k}'h\uparrow}
  \alpha_{{\bf k}h\uparrow}   \alpha_{{\bf k}e\uparrow},\label{Uc}
\end{eqnarray}
where $U_{{\bf k}{\bf k}'} =(U_{{\bf k}{\bf k}'}^{ab}
e^{i(\theta'-\theta)} +U_{{\bf k}'{\bf k}}^{ab}
e^{i(\theta-\theta')} +U_{{\bf k}{\bf k}'}^{aa}+U_{{\bf k}{\bf
k}'}^{bb}
 )/4$ with $U_{{\bf k}{\bf k}'}^{ab} =\sum_j U^{ab}_{0j} e^{-i({\bf
 k}-{\bf k}')\cdot ({\bf r}_j+{\bf \delta})}$
and $U_{{\bf k}{\bf k}'}^{ss} =\sum_j U^{ss}_{0j} e^{-i({\bf
 k}-{\bf k}')\cdot {\bf r}_j}$ for the coordinate ${\bf r}_j$ of
the site $j$ and the lattice spacing vector ${\bf \delta}$. $U_{\bf
k k'}$ is a large positive value at ${\bf k}={\bf k'}$ and it
gradually and oscillatorily decays to zero with increase of $|{\bf
k}-{\bf k'}|$. As discussed before, this attractive interaction does
not induce the e-h recombination while it binds the electrons and
holes into pairs. The mean field approximation of Eq. (\ref{Uc})
reads
\[
U_C\approx \sum_{\bf k} \Delta_{\bf k} \alpha^{\dagger}_{{\bf
k}e\uparrow} \alpha^{\dagger}_{{\bf k}h\uparrow} + \sum_{\bf k}
\Delta_{\bf k}^* \alpha_{{\bf k}h\uparrow} \alpha_{{\bf
k}e\uparrow}\] with the e-h pair condensation order parameter
$\Delta_{\bf k} \equiv
$$-\sum_{{\bf k}'} U_{{\bf k}'{\bf k}} \langle \alpha_{{\bf
k}'h\uparrow} \alpha_{{\bf k}'e\uparrow} \rangle$.

Comparing with the spin singlet Cooper pair with charge $2e$, this
e-h pair is of spin $\hbar$ and charge neutral. The total mean field
Hamiltonian then is given by
\begin{eqnarray}
H_{MF}=\sum\limits_{\bf k}( \alpha^{\dagger}_{{\bf k}e\uparrow},
\alpha_{{\bf k}h\uparrow}
    )
    \left(\begin{array}{ll} \epsilon_{+\uparrow} & \Delta_{\bf k} \\
    \Delta_{\bf k}^* & \epsilon_{-\downarrow} \end{array}\right)
 \left( \begin{array}{l}
 \alpha_{{\bf k}e\uparrow} \\ \alpha^{\dagger}_{{\bf k}h\uparrow}
 \end{array}
 \right).
\end{eqnarray}
The energy spectrum for the mean field Hamiltonian $H_{MF}$ is
shown in Fig. \ref{fig:Fig.1}(d). An energy gap with the
magnitude of $|\Delta_{\bf k}|$ is opened. When an electron and a
hole combine into an e-h pair, the energy of the system is
reduced by $2|\Delta_{\bf k}|$. This means the condensed state of
the e-h pairs is more stable than the unpaired one. Thus, the
ground state of FM graphene is a neutral superfluid with spin
$\hbar$ per pair, namely, a SSC state. The spin current can
dissipationlessly flow in the SSC and its spin resistance is zero.

The energy gap $\Delta$ can be estimated as follows. By using the
definition  $\Delta_{\bf k} \equiv -\sum_{{\bf k}'} U_{{\bf k}'{\bf
k}} \langle \alpha_{{\bf k}'h\uparrow} \alpha_{{\bf k}'e\uparrow}
\rangle$ and the Hamiltonian $H_{MF}$, one has the self-consistent
equation:
\begin{equation}
 \Delta_{\bf k} =\sum\limits_{{\bf k}'} (U_{{\bf k}{\bf k}'}
  \Delta_{{\bf k}'}/2A) \{f(-A)-f(A) \},\label{self}
\end{equation}
where $f(A) =1/[\exp(A/k_B T) +1]$, $A=\sqrt{(M-k')^2 +\Delta_{{\bf
k}'}}$ and $T$ is the temperature. At zero temperature and assuming
$U_{{\bf k}{\bf k}'}=U\theta(k_D-|{\bf k}-{\bf k'}|)$ with the
cut-off momentum $k_D$, the self-consistent equation (\ref{self})
reduces to $1 =(U/2)\sum_{\bf k}
\theta(k_D-k)/\sqrt{(M-k)^2+\Delta^2} = \frac{\sqrt{3}U}{3
t^2}\int_0^{\epsilon_D} d\epsilon_k \frac{2\pi
\epsilon_k}{\sqrt{(M-\epsilon_k)^2+\Delta^2}}$, where
$\epsilon_D=v_F k_D$. Numerically, we solve the self-consistent
equation by using the e-e interaction $U_C$ with the nearest
neighbor cut-off. The gaps vary as the cut-off $\epsilon_D$ for a
fixed $M$ or as the FM split energy $M$ for a fixed $\epsilon_D$ are
shown in Fig. \ref{fig:Fig.1}(e) and (f), respectively. We see that
the gap $\Delta$ grows faster than an exponential function with
increase of $\epsilon_D$. When $M=5meV$ and
$\epsilon_D=0.18t$,\cite{ref5,ref6} one gets $\Delta\approx 3$meV.
This yields the critical temperature $T_C$ of the transition from
the normal state to the SSC at about $30K$. Similar to the case in a
superconductor, in the presence of weak impurities, $T_C$ is
slightly reduced but the SSC phase can still exists if $T<T_C$,
except in the case when the impurity strength is larger than
$\Delta$ and its density is higher than $1/\xi^2$ with $\xi$ being
the coherence length $\xi=\hbar v_F/\Delta$.

\begin{figure}
\includegraphics[width=8cm,totalheight=6cm]{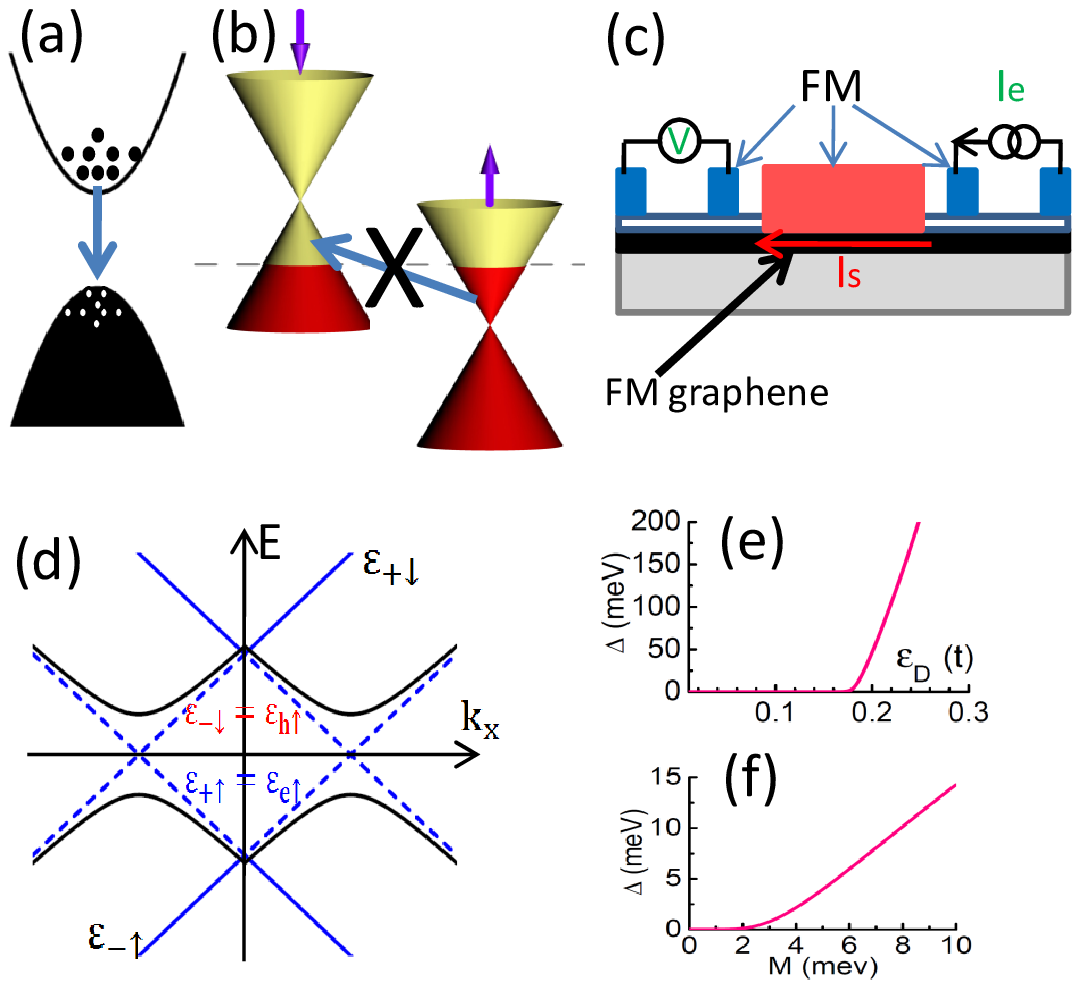}
\caption{\label{fig:Fig.1} (color online). The schematic
diagrams: (a) the band structure for the conventional exciton
system; (b) for FM graphene; and (c) for the proposed
four-terminal device to measure the SSC state. (d) The schematic
energy bands of FM graphene for free electrons (blue curves) and
for the e-h pair condensate (black curves). (e) The gap $\Delta$
vs. $\epsilon_D$ at the FM magnetic moment $M=5meV$ and (f) The
gap vs. $M$ at $\epsilon_D=0.18t$. } \label{fig:1}
\end{figure}

Meissner effect is the criterion that a superconductor differs from
a perfect metal: The magnetic field can not enter the bulk of a
superconductor.\cite{meiss} This phenomenon can be described by the
London equations.\cite{london} Is there a Meissner-like effect for
the SSC? Consider a SSC with the superfluid carrier density $n_s$ in
an electric field ${\bf E}$ and a magnetic field ${\bf B}$. A
magnetic force ${\bf F}=({\bf m}\cdot \nabla){\bf B}$ acts on these
spin carriers. Here ${\bf m}=(4\pi g\mu_B/h) {\bf s}$ is the
magnetic moment of a carrier; $\mu_B$ is the Bohr magneton and $g$
is the Lande factor. This force accelerates the carrier by Newton's
second law ${\bf F}= m^* d {\bf v}/dt$ for a carrier with the
velocity $ {\bf v}$ and the effective mass $m^*$. The spin current
density $\mathbbm{J}_s =n_s {\bf v} {\bf s}$ is thus a tensor. The
time derivative of this super-spin-current density $\mathbbm{J}_s$
is then given by
\begin{equation}
d\mathbbm{J}_s/dt = a(\mathbf{s}\cdot \nabla)
\mathbf{B}\mathbf{s},\label{7}
\end{equation}
with the constant $a =4\pi g\mu_B n_s/hm^*$. Comparing with London's
first equation for the super-charge-current density:\cite{london}
$d{\bf J}/dt\propto {\bf E}$, the spatial variation of ${\bf B}$
along with the magnetic moment plays the role of an external field
accelerating the spin carriers.

When an electric field ${\bf E}$ applies, by acting $({\bf s}\cdot
\nabla)$ on two sides of the Maxwell equations$\nabla \times {\bf B}
=\mu_0\epsilon_0
\partial {\bf E} /\partial t$ and using Eq.(\ref{7}), we obtain
$\frac{\partial}{\partial t} [\nabla \times \mathbb{J}_s ] =
\frac{\partial}{\partial t} [\mu_0\epsilon_0 a ({\bf s} \cdot
\nabla) {\bf E}{\bf s} ]$. Integrating over the time $t$, one has
the equation for $\mathbb{J}_s$
\begin{equation}
\nabla \times \mathbb{J}_s  = \mu_0\epsilon_0 a ({\bf s} \cdot
\nabla) {\bf E}{\bf s},\label{8}
\end{equation}
where the integral constant is taken to be zero because of the
requirement of thermodynamic equilibrium. Instead of the magnetic
field in London's second equation for the
superconductor,\cite{london} the `external field' here is the
spatial variation of the electric field ${\bf E}$ along with the
magnetic moment.

Eqs. (\ref{7}) and (\ref{8}) for $\mathbb{J}_s$ play roles similar
to the London equations in superconductor.\cite{london} For example,
if the system is in the steady state, $d\mathbb{J}_s/dt = a({\bf
s}\cdot \nabla) {\bf B}{\bf s} =0$ implies that the variation $({\bf
s}\cdot \nabla) {\bf B}$ of the magnetic field along the direction
of the FM magnetic moment ${\bf m}$ must be zero because of the zero
spin resistance. On the other hand, Eq.(\ref{8}) means the variation
of the electric field, $({\bf s} \cdot \nabla) {\bf E}$, is zero in
bulk of the SSC. This is an electric `Meissner effect' in the SSC
against a spatial variation of an electric field.

\begin{figure}
\includegraphics[width=7.5cm,totalheight=3cm]{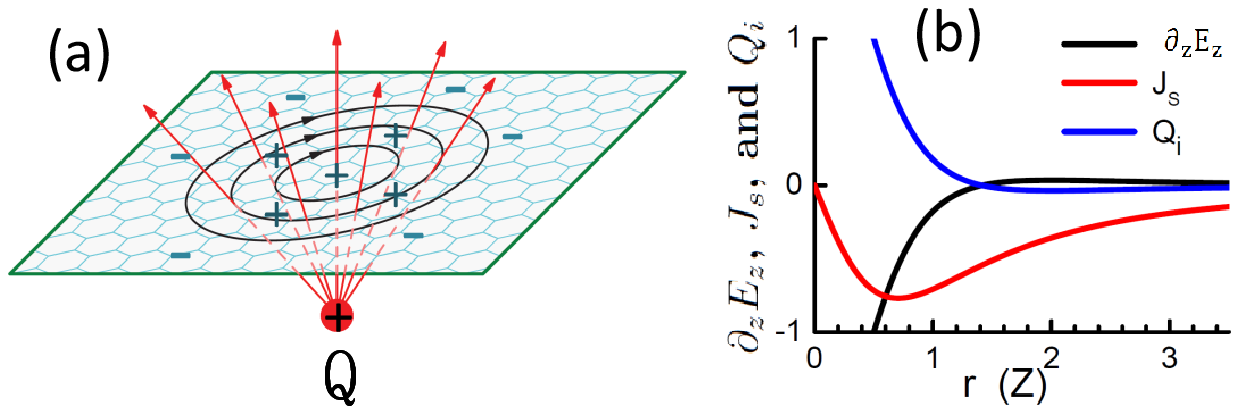}
\caption{\label{fig:Fig.2} (color online). (a) The schematic
diagram for the device consisting of a positive charge $Q$ and FM
graphene. (b) The variation $\partial_z E_z$ ($Q/4\pi\epsilon_0$)
of the electric field, the induced super-spin-current $J_s$
($\mu_0 a Q/8\pi$), and the equivalent charge $Q_i$ (i.e.
$\partial_z E_z^i$) versus the radial distance $r$. }
\label{fig:2}
\end{figure}

We now give an example of this electric Meissner effect. Consider
a positive charge $Q$ at the origin and an infinite FM graphene
in the $x$-$y$ plane at $z=Z$ as shown in Fig.
\ref{fig:Fig.2}(a). The charge $Q$ generates an electric field
${\bf E}$ in  FM graphene plane. This electric field will induce
a super-spin-current in graphene against the spatial variation of
${\bf E}$. Assuming that the magnetic moment ${\bf m}$ (i.e.,
${\bf s}$) is in the $z$-direction, then $({\bf s} \cdot \nabla)
E_z =\partial_z E_z = \frac{Q}{4\pi \epsilon_0}
\frac{r^2-2z^2}{(z^2+r^2)^{5/2}}$ with $r^2=x^2+y^2$. Solving
Eq.(\ref{8}), one has the induced super-spin-current density
$J_s=-\frac{\mu_0 a Q}{4\pi} \frac{r}{(Z^2+r^2)^{3/2}}$. This
$J_s$ flows along the tangential direction (see
Fig.\ref{fig:Fig.2}(a)) and its spin points to the $z$-direction.
On the other hand, as a usual spin current,\cite{ref9} the
super-spin-current density $J_s$ can generate an electric field
${\bf E}^i$ in space which is the same as that generated by the
electric dipole moment $\vec{p}_e \propto
(-\frac{r}{(Z^2+r^2)^{3/2}}, 0,0 )$ or the equivalent charge $Q_i
= -\nabla \cdot \vec{p}_e \propto
\frac{2Z^2-r^2}{(Z^2+r^2)^{5/2}}$. In Fig.\ref{fig:Fig.2}(b), we
plot the radial distributions in graphene for the variation
$\partial_z E_z$ of the electric field of the original charge
$Q$, the induced super spin current density $J_s$, and the
equivalent charge $Q_i$ (i.e. the spatial variation $\partial_z
E^{i}_z$). For $r<\sqrt{2} Z$ ($r>\sqrt{2} Z$) with $\partial_z
E_z$ being negative (positive), the spatial variation $\partial_z
E^{i}_z$ of the electric field ${\bf E}^{i}$ induced by the
super-spin-current is positive (negative). As a result,
$\partial_z E^{i}_z$ counteracts the variation $\partial_z E_z$
and then causes the variation of the total electric field in the
SSC to vanish.

\begin{figure}
\includegraphics[width=8cm,totalheight=5.5cm]{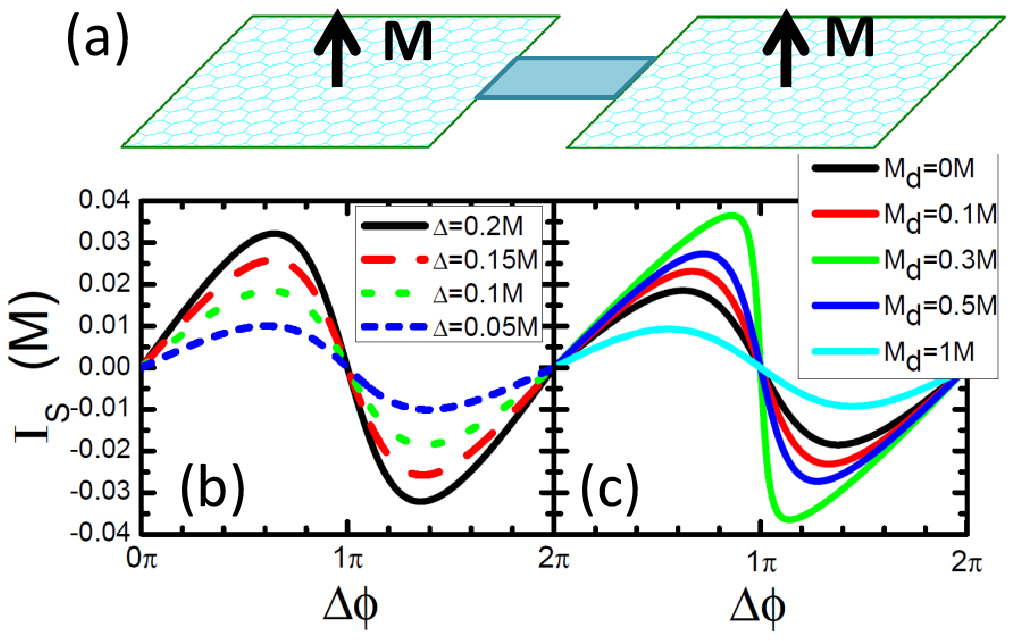}
\caption{\label{fig:Fig.3} (color online). (a) The schematic
diagram of the weakly coupled SSC-normal conductor-SSC junction.
(b) The spin current $I_s$ vs. the phase difference $\Delta\phi$
for different $\Delta$ and $M_d=0$. (c) The $I_s$-$\Delta\phi$
curves for different $m_d$ and $\Delta=0.1M$. Here $\epsilon_d=0$
and $\Gamma\equiv 2\pi t^2_{\beta}\rho_{k} = 0.1 M$; $\rho_k$ is
the density of state of FM graphene in momentum space.}
\end{figure}

Josephson effect is another highlight of superconductivity and has wide
applications.\cite{jose} We now investigate the similar effect for
this SSC by considering a device consisting of two SSCs which are
weakly coupled by a normal conductor, i.e., a SSC/normal
conductor/SSC junction (see Fig. \ref{fig:Fig.3}(a)). We can
explicitly show the existence of the super-spin-current in this
device in equilibrium. The weakly coupled junction is described by
the Hamiltonian $H=\sum_{\beta (\beta=L,R)} H_{\beta} + H_c + H_T$,
where
\begin{eqnarray}
H_{\beta} & = & \sum\limits_{\bf k}
 ( \alpha^{\dagger}_{\beta{\bf
 k}e\uparrow}, \alpha_{\beta{\bf k}h\uparrow}
    )
    \left(\begin{array}{ll} \epsilon_{+\uparrow} & \Delta_{\beta{\bf k}} \\
    \Delta_{\beta{\bf k}}^* & \epsilon_{-\downarrow} \end{array}\right)
 \left( \begin{array}{l}
 \alpha_{\beta{\bf k}e\uparrow} \\ \alpha^{\dagger}_{\beta{\bf k}h\uparrow}
 \end{array}
 \right ), \nonumber\\
H_c &=& \sum\limits_{\sigma} (\epsilon_d +\sigma M_d)
c^{\dagger}_{\sigma} c_{\sigma} , \nonumber\\
H_T &=& \sum\limits_{\beta,{\bf k}} \left[ t_{\beta}
  \alpha^{\dagger}_{\beta {\bf k} +\uparrow} c_{\uparrow}
 + t_{\beta} \alpha^{\dagger}_{\beta {\bf k} -\downarrow} c_{\downarrow}
 +H.c.\right]. \nonumber
\end{eqnarray}
Namely, $H_{L/R}$, $H_c$, and $H_T$ are the Hamiltonians of the
left/right SSC, the normal conductor, and the tunnelings between
them, respectively. The order parameters $\Delta_{L/R {\bf k}}
=\Delta e^{i\phi_{L/R}}$, where the SSC phases $\phi_{L/R}$ are
assumed to be independent of the momentum ${\bf k}$. We consider a
phase difference $\Delta\phi\equiv\phi_L-\phi_R$ between the left
and right SSCs, which origins from a spin current flowing through
the junction under the drive of an external device or from a
variation of an external electric field thread the ring junction
device. The normal conductor is described by a level (or a quantum
dot) with the spin index $\sigma$ and spin-split energy $M_d$.

The spin-dependent particle current $I_{\beta\sigma}$ with the
spin $\sigma$ from the $\beta$ SSC to the central normal
conductor can be calculated by the following equation,\cite{ref8}
$I_{\beta\sigma} = Re (2t^*_{\beta}/\hbar) \int
\frac{d\epsilon}{2\pi} G^<_{\beta\sigma,\sigma}(\epsilon)$, where
the lesser Green function $G^<_{\beta\sigma,\sigma}(\epsilon)$ is
the Fourier transformation of $G^<_{\beta\sigma,\sigma}(t)\equiv
i\langle c^{\dagger}_{\sigma}(0) \alpha_{\beta {\bf k} \pm
\sigma}(t)\rangle$. This Green function
$G^<_{\beta\sigma,\sigma}(\epsilon)$ can be calculated by using
the Dyson equation and so is the particle current
$I_{\beta\sigma}$.\cite{ref8} Therefore, one can obtain the spin
current $I_s = (I_{R\uparrow}-I_{R\downarrow})\hbar/2$ and the
charge current $I_e = (I_{R\uparrow}+I_{R\downarrow})e$. The
charge current $I_e$ is identically zero because the e-h pairs
are charge neutral. The spin current $I_s$ versus $\Delta\phi$ in
the equilibrium with zero bias and zero spin bias is calculated
and shown in Fig. \ref{fig:Fig.3}. There is a super-spin-current
flowing through the junction that resembles the Josephson
tunneling in a conventional superconductor junction. While Fig.
\ref{fig:Fig.3}(b) exhibits the $I_s$-$\Delta\phi$ curves for
different values of the gap $\Delta$ with $m_d=0$, Fig.
\ref{fig:Fig.3}(c) shows $I_s$ can also be observed in non-zero
$M_d$ as long as $\Delta \ne 0$ and $\Delta\phi\ne 0,\pi$.

In conclusion, a SSC state as well as the electric `Meissner
effect' and spin-current Josephson effect are predicted in FM
graphene. For detection of the SSC, one can measure the zero spin
resistance or super spin current. Based on the experiment in
Ref.\cite{ref1b}, here we propose a four-terminal device (as
shown in Fig.1c) which can be used to measure the non-local
resistance and then confirm the SSC state.\cite{sm} A non-local
resistance measurement means that a current is applied to the two
right electrodes and the bias is measured on the two left
electrodes.\cite{ref1b} In this device, the non-local resistance
is induced by a pure spin current flowing from right side through
FM graphene (SSC) to left side.\cite{ref1b} Once FM graphene
turns into the SSC state by lowering temperature $T<T_c$, the
spin current will flow with no dissipation, leading to a sharp
increase of the non-local resistance (see the Fig.1c in the
supplemental material). Due to the zero spin resistance effect in
SSC, this sharp increase can be observed even in a macroscopic
SSC device.

{\bf Acknowledgments:} This work was financially supported by
NSF-China under Grants Nos. 10734110, 10974015, 11074174, and
10874191, China-973 program and US-DOE under Grants No. DE-FG02-
04ER46124.

\hspace{8mm}

{\bf Supplementary information:}

In this supplementary material we discuss how to detect the spin
superconductor state. In the following we first suggest five
measurable physical quantities or methods, and then follow up by
proposing an experimental setup.

1) When the system enters the spin superconductor state, an energy
gap opens up (see Fig.1d in the paper). This energy gap can be
measured by ARPES or STM. When the temperature $T$ is lower (or
higher) than the critical temperature $T_C$, the gap emerges (or
disappears).

2) Because of the opening of an energy gap and the spin
superconductor state does not carry charge current, the resistance
sharply increases when the FM graphene enters from the normal metal
state to the spin superconductor state. This sharp increase in
resistance can be experimentally tested.

Notice that the measurements 1) and 2) are commonly done. These
measurements demonstrate the opening of an gap and give strong hint
that the sample may enter into the spin superconductor.

3) The zero spin resistance is a main characteristic for the spin
superconductor. Due to the zero spin resistance, the spin current
can flow without any dissipation even for a macroscopic sample. At
the end of the supplementary material we propose a four-terminal
device and in this device one can show zero spin resistance by
measuring a non-local resistance.

4) In the Meissner effect for spin superconductor, an electric field
applied to the spin superconductor can induce the super spin current
on the surface of the sample. This induced super spin current can
generate an electric field ${\bf E}^i$ that against the variation of
the external electric field. Here the induced electric field ${\bf
E}^i$ is equivalent to that generated by a certain surface charge
distribution. In this case, the induced electric field and the
equivalent surface charge distribution are the measurable
quantities. For the example of Fig.2 in the paper, the equivalent
surface charge $Q_i$ is proportional to $\frac{n_s}{m^*}
\frac{2Z^2-r^2}{(Z^2+r^2)^{5/2}}$, this value is quite large due to
the smallness of the effective mass $m^*$ in graphene. But the
equivalent surface charge $Q_i$ is still finite even if $m^*=0$,
since once it reaches the complete shielding and the variation of
the total electric field is zero, no more super spin current and
charge are induced. By considering the complete shielding case, the
inducing surface charge $Q_i =\frac{2Z^2-r^2}{(Z^2+r^2)^{5/2}}
\frac{Qd}{4\pi}$, where $d\approx 0.1 nm$ is the thickness of
graphene. Let us estimate the value of $Q_i$. We assume that the
distance $Z$ between the external charge $Q$ and the plane of
graphene is $10nm$ and $Q$ is a basic charge. The inducing surface
charge $Q_i$ is about $10^{13}m^{-2}$ at the position $r=0$. This
surface charge is quite large and should be measurable.

5) The super spin current in the the spin superconductor state or in
the spin-current Josephson effect is also a measurable physical
quantity. For example, it can be directly measured by observing the
second-harmonic generation of the Faraday rotation as done in
reference, Nature physics 6, 875 (2010).

In addition, many indirect methods have successfully  measured the
spin current, e.g., by probing the spin accumulation (Science 306,
1910(2004)), or probing the bias due to the inverse spin Hall effect
(Nature 442, 176(2006)), or probing the bias through a quantum point
contact (Nature 458, 868(2009), etc. These methods can also be used
in our system. Let us imagine a spin current flowing into the spin
superconductor (FM graphene) from one terminal, this spin current
can flow through the spin superconductor with no dissipation and
flows out to another terminal. Then we can measure the (usual) spin
current at the outside through the aforementioned methods.

Notice that it is not necessary to do all of the aforementioned
measurements. In fact, if one can take a measurement either in 3),
4), or 5), it establishes the presence of the spin superconductor.

{\bf Now let us suggest an experiment setup and propose a detailed
experimental process.}

Based on the device in the paper of Nature 488, 571 (2007), we
propose a four-terminal device consisting of a graphene ribbon
coupled by four FM electrodes and a long FM strip, as shown in the
Fig.1(a) and (b). Except for the long FM strip (the red region), the
device is what used in the paper of Nature 488, 571 (2007). So this
device can be realized by the present technology. The ideal case is
that the long red strip is a FM insulator which directly couples to
the graphene without the $Al_2 O_3$ layer. But even if this is hard
to achieve, it is also fine for the red region being a FM metal.
Graphene in the red region has a FM exchange splitting, and it is
normal FM graphene at high temperature and turns to the spin
superconductor at low temperature. Now we can qualitatively analyze
the measurement results if the spin superconductor is realized. To
simplify the analysis, we only consider the magnetic moments in all
FM electrodes and strip are in the same direction.

\begin{figure}
\includegraphics[width=7cm,totalheight=7cm]{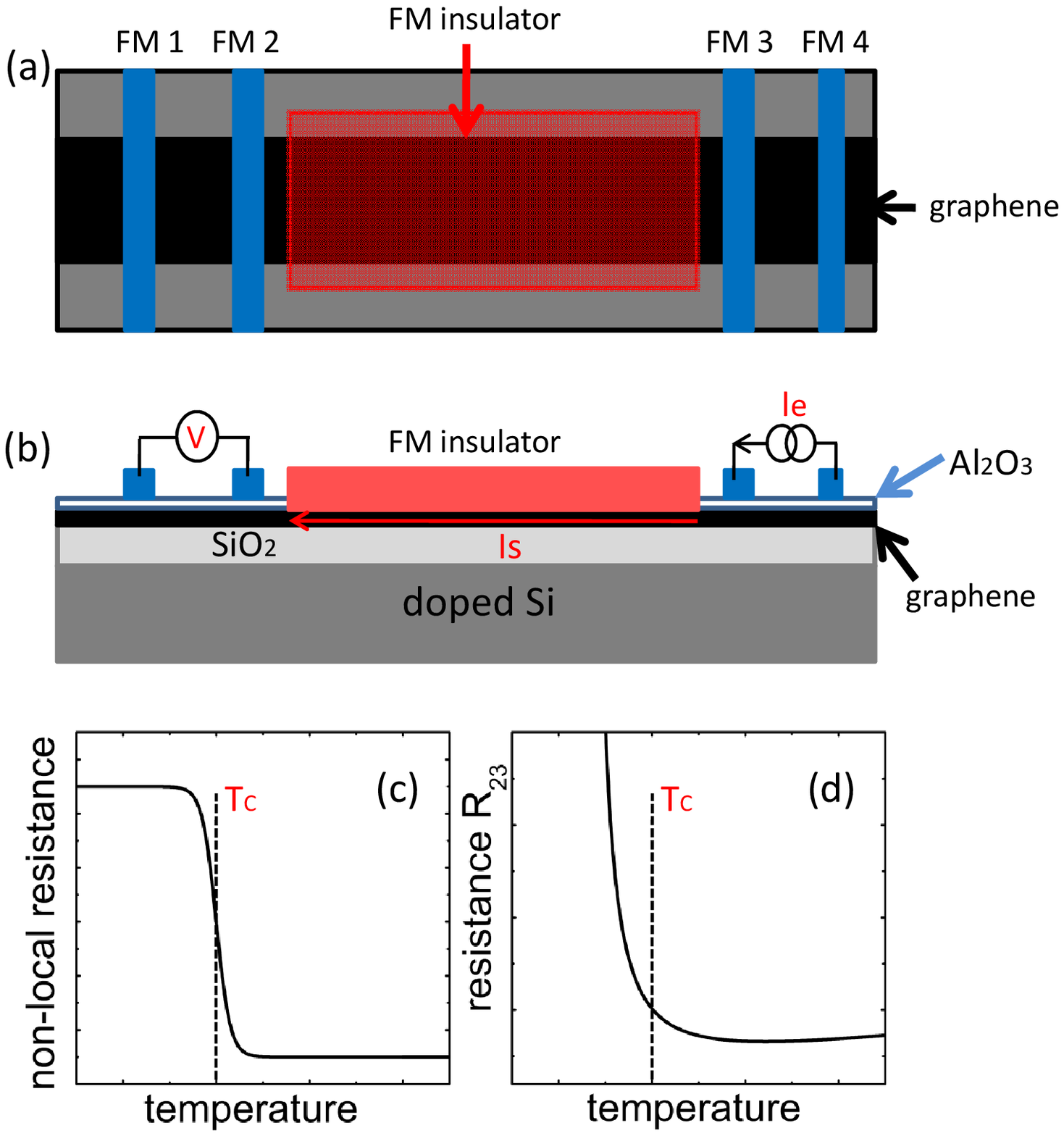}
\caption{\label{fig:Fig.1} (color online). (a) and (b) are schematic
diagrams for the proposed device. (a) is the top view and (b) is the
side view. (c) and (d) schematically show the non-local resistance
and the resistance $R_{23}$ versus the temperature, respectively. }
\label{fig:1}
\end{figure}

Here we measure the non-local resistance as in the experimental
paper of Nature 488, 571 (2007). In this measurement, a current is
applied to two right electrodes (electrodes 3 and 4) and the bias is
measured on two left electrodes (electrodes 1 and 2). When a current
is applied between electrodes 3 and 4, it injects a pure spin
current into graphene. This pure spin current flows from the
electrode 3 through FM graphene (spin superconductor) to the left
side (as shown in the Fig.1(b)). Then it induces the bias between
the electrodes 1 and 2. When FM graphene is in the normal FM metal
phase, it has a finite spin resistance. In this case, the spin
current gradually decays along its transport direction, so the bias
and the non-local resistance $R_{12,34}$ are small.  Denoting
$L_{ij}$ the distances between the electrode $i$ and $j$ and
assuming $L_{12}$ and $L_{34}$ are much longer than the spin
relaxation length $\lambda$ of the graphene in the normal state, the
non-local resistance $R_{12,34}$ can
 be obtained analytically (see Nature 416, 713(2002); Nature 442,
176(2006); etc): $R^{normal}_{12,34}= \frac{1}{2} P^2 \frac{\lambda
}{\sigma_{G} W} e^{-L_{23}/\lambda}$, where $W$ is the width of the
graphene ribbon, $P$ the spin polarization of the FM electrode, and
$\sigma_{G}$ the conductivity of normal FM graphene. On the other
hand, when FM graphene is in the spin superconductor phase, the spin
resistance is zero and the spin current can flow through it with no
dissipation. In this case, the non-local resistance can be obtained
as $R_{12,34}^{SSC} = \frac{1}{2} P^2 \frac{\lambda }{\sigma_{G}W}
e^{-(L_{23}-L_{SSC})/\lambda} = R^{normal}_{12,34}
e^{L_{SSC}/\lambda}$, which are much larger than
$R^{normal}_{12,34}$. Therefore, a sharp increase in the non-local
resistance can be observed (see the Fig.1(c)) when  FM graphene
turns from the normal metal phase to the spin superconductor phase
as temperature is lowered. Furthermore, the spin resistance in the
spin superconductor is always zero regardless of the length of the
spin-superconductor device, so the non-local resistance is
independent of the length of spin superconductor system and the
sharp increase in non-local resistance can be observed even in a
macroscopic spin superconductor device. Experimentally, one can also
observe the relation between the non-local resistance and the sample
length to confirm the zero spin resistance.

In addition, if the long red strip is a FM insulator, a sharp
increase can also be observed in the resistance $R_{23}$ between
electrodes 2 and 3 (see the Fig.1(d)), as discussed in point 2).

Finally, we emphasize that the main body of the proposed device as
well the pure spin current injected into graphene have been realized
in an experiment (see Nature 488, 571 (2007), etc). So the proposed
method is experimentally feasible.


\end{document}